\documentstyle[twocolumn,epsf]{jpsj}
\def\runtitle{ 
 Correlation Effects in  
 a One-Dimensional Quarter-Filled Electron System 
  with Repulsive Interactions

}
\def\runauthor{Hideo {\sc Yoshioka}, Masahisa {\sc Tsuchiizu} 
                 and Yoshikazu {\sc Suzumura}}
\def\d{{\rm d}}
\def\lan{\left\langle}
\def\ran{\right\rangle}

\def\e{{\rm e}}
\def\virg{\;\;,}
\def\point{\;\,.}
\def\vf{v_{\rm F}}
\def\kf{k_{\rm F}}

\def\ggs{\buildrel\textstyle > \over {\hbox{\raise0.2ex\hbox{$\sim$}}}}
\def\lls{\buildrel\textstyle < \over {\hbox{\raise0.2ex\hbox{$\sim$}}}}
\def\gsim{\,\lower0.75ex\hbox{$\ggs$}\,}
\def\lsim{\,\lower0.75ex\hbox{$\lls$}\,}
\def\N{\hat{N}}

\def\im{{\rm i}}
\def\ie{{\it i.e.}, }
\def\delx{\partial_x}
\def\jo #1#2#3#4{#1 {\bf #2} (#3) #4} 

\def\PRB{Phys.\ Rev.\ B}
\def\PRL{Phys.\ Rev.\ Lett.}

\def\JPSJ{J.\ Phys.\ Soc.\ Jpn.}

\def\ADV{Adv.\ Phys.}

\def\EPL{Europhys.\ Lett}

\title
{
 Correlation Effects in a  
 One-Dimensional Quarter-Filled Electron System 
  with Repulsive Interactions 
}

\author{
Hideo {\sc Yoshioka}$^a$\hspace{-0.5mm}\footnote{E-mail: h44770a@nucc.cc.nagoya-u.ac.jp}\hspace{0.5mm}, 
Masahisa {\sc Tsuchiizu}$^a$
   and Yoshikazu  {\sc Suzumura}$^{a,b}$
}

\inst{
$^a$Department of Physics, Nagoya University, Nagoya 464-8602\\
$^b$CREST, Japan Science and Technology Corporation (JST)
}

\recdate{December 6, 1999}

\abst{ 
 A one-dimensional electron system at quarter-filling 
  has been examined by applying the renormalization group method 
    to a bosonized model   with on-site ($U$) and 
      nearest-neighbor ($V$)  repulsive interactions.  
  By evaluating  both normal scattering and Umklapp scattering 
   perturbatively, we obtain a phase diagram in which 
 a metallic state with  a $2\kf$ spin density wave 
    ($\kf$ is the Fermi wave number) moves into  
     an insulating state  with charge disproportionation of  
       a $4\kf$ charge density wave 
 with an increase in both $U$ and $V$.
 The effect of the next-nearest-neighbor repulsion 
  is also discussed. 
}

\kword{extended Hubbard model, quarter-filling, commensurability energy, 
 spin density wave, charge density wave
}

\begin{document}
\sloppy
\maketitle

A one-dimensional (1-D) electron system at quarter-filling is a basic model 
  for understanding the electronic properties of quasi 1-D organic
     conductors.\cite{review}  
The 1-D electron system with only on-site Coulomb repulsive interaction,
   $U$ (1-D Hubbard model), 
    does not exhibit a  metal-insulator (M-I) transition 
   as a function of $U$, and is  metallic at any filling, except 
     half-filling.\cite{Lieb-Wu} 
The most dominant state is given by $2\kf$ spin density wave
 ($2\kf$-SDW),  where $\kf$ is the Fermi wave number. 
 The presence of  the long-range repulsive interaction is expected 
 to enrich the phase at quarter-filling.
 In fact,  numerical diagonalization for the model with 
   both $U$ and  the nearest-neighbor interaction, $V$,
\cite{Mila-Zotos,Penc-Mila-I,Sano-Ono,Nakamura-Kitazawa-Nomura}       
 shows that 
 the insulating phase appears for a large strength of both $U$ and $V$  
 and that the superconductivity becomes  the most dominant fluctuation
   for a large $V$ and small $U$.
 On the other hand,  several properties have been elucidated  within 
    the mean-field theory.
\cite{Seo-Fukuyama,Suzumura,Kobayashi-Ogata-Yonemitsu}
   With increasing $V$,  a  transition occurs 
 from   a pure $2\kf$-SDW state to 
   a  coexistent state of $2\kf$-SDW and 
     $4\kf$ charge density wave ($4\kf$-CDW).
\cite{Seo-Fukuyama} 
 Such a coexistence is  maintained for   
   the  charge ordering in (DI-DCNQI)$_2$Ag observed  
   by the  $^{13}$C-NMR measurement.\cite{Hiraki-Kanoda}
 The transition has been examined 
  by evaluating  
     the commensurability energy
     corresponding to the $8\kf$-Umklapp 
       scattering.\cite{Suzumura}
 The next-nearest-neighbor repulsion  results in  
    a coexistence of $2\kf$-SDW and purely electronic $2\kf$-CDW,
\cite{Kobayashi-Ogata-Yonemitsu} and it 
 has been proposed to be the origin of the coexistence   
 observed by  the X-ray experiment on  
(TMTSF)$_2$PF$_6$.
\cite{Pouget-Ravy}

 At quarter-filling, 
 the $8\kf$-Umklapp scattering 
 is crucial to obtain the commensurability energy. 
 Although the above mean-field results may show common features,
   we need to calculate the Umklapp scattering in the 1-D system   
    by taking into account quantum fluctuation. 
 The existence of  
   commensurability energy of such high order 
  has been pointed out.\cite{Giamarchi-Millis,Schulz,Giamarchi,Yonemitsu}  
However,  to the best of our knowledge, 
 there are no studies for the phase diagram 
 on the plane of $U$ and $V$, which is calculated 
   by  using the analytical expression 
    of the commensurability energy.
    
 In the present paper, 
  a 1-D system with repulsive interactions at quarter-filling 
  is investigated
   using  the bosonization method and 
    the renormalization group (RG) theory. 
 Based on 
 the commensurability energy,
\cite{Tsuchiizu} 
 a  phase diagram   
 is derived on the plane of $U$ and $V$.  
The relevance of the present results 
 to  the observation in (DI-DCNQI)$_2$Ag salt,
\cite{Hiraki-Kanoda}
 and 
  the effect of the next-nearest-neighbor interaction
are discussed.

We consider 
 a 1-D  extended Hubbard model  
   given by the Hamiltonian, ${\cal H} = {\cal H}_0 + {\cal H}_{\rm int}$,    
\begin{eqnarray}
{\cal H}_0 
&=& - t \sum_{j \sigma} 
   \left( a_{j, \sigma}^\dagger  a_{j+1, \sigma}  + {\rm h.c.}\right) 
   -\mu \sum_{j, \sigma} n_{j, \sigma}  \nonumber \\
&=& \sum_{K \sigma} (\epsilon_K - \mu ) a^\dagger_{K, \sigma} a_{K, \sigma}     \label{eqn:H0} \virg \\
{\cal H}_{\rm int} &=& \frac{U}{2} \sum_{j \sigma} 
n_{j, \sigma} n_{j, -\sigma} + 
V \sum_{j \sigma \sigma'} n_{j, \sigma} n_{j+1, \sigma'}  \nonumber \\
&=& \frac{1}{N_L} \sum_{\sigma \sigma'} \sum_{K_1 \sim K_4}
\left\{
\frac{U}{2} \delta_{\sigma, -\sigma'} + V {\rm e}^{- \im (K_2 - K_3)a}
\right\} \nonumber \\
&\times& \delta_{K_1 + K_2 - K_3 - K_4, G}
a^\dagger_{K_1, \sigma} a^\dagger_{K_2, \sigma'} a_{K_3, \sigma'} a_{K_4, \sigma} \virg
\end{eqnarray}
where $t$ and $\mu$ denote 
  the transfer energy  and   chemical potential, respectively, and  
 $\epsilon_K = -2t \cos Ka$ with   lattice constant $a$ and  
$- \pi/a < K \leq \pi/a$.
The quantity 
$a_{j, \sigma}^\dagger ( = 1/\sqrt{N_L} \sum_K \e^{-\im K a j} 
a_{K, \sigma}^\dagger) $ denotes the creation
operator of the electron at the $j$-th site with spin
$\sigma$, $n_{j, \sigma} = a_{j, \sigma}^\dagger a_{j, \sigma}$, 
  $G = 0, \pm 2 \pi/a$, and 
 $N_L$ is the number of the lattice. 
 For the later convenience, we divide 
the one-particle states as 
$d_{k, -, \sigma} = a_{K, \sigma}$ for $- \pi /a < K \leq - \pi / (2a)$, 
$c_{k, -, \sigma} = a_{K, \sigma}$ for $- \pi /(2a) < K \leq 0$,
$c_{k, +, \sigma} = a_{K, \sigma}$ for $0 < K \leq \pi/(2a)$, and
$d_{k, +, \sigma} = a_{K, \sigma}$ for $\pi /(2a) < K \leq \pi / a$,
$k$ being the deviation of the wave number from $\pm \kf$ for $c_{k, \pm,
\sigma}$, and $\pm 3\kf$ for $d_{k, \pm, \sigma}$.
\cite{Penc-Mila-II}  
In terms of $c_{k, p, \sigma}$ and $d_{k, p, \sigma}$, 
${\cal H}_0$ is written as 
${\cal H}_0 = \sum_{p k \sigma} 
\left\{
(\epsilon_{p\kf + k} - \mu ) c^\dagger_{k, p, \sigma} c_{k, p, \sigma}
+ (\epsilon_{3p\kf + k} - \mu ) d^\dagger_{k, p, \sigma} d_{k, p, \sigma}
\right\} $
 and 
 ${\cal H}_{\rm int}$
is  rewritten as 
 ${\cal H}_{\rm int} = \sum_{i=0}^4 {\cal H}_{{\rm int},i}$,  
 where $i$ denotes  
 the number of $d_{k, p, \sigma}$ and/or $d^{\dagger}_{k, p, \sigma}$ 
 in the respective interactions. 
  
When  the one-particle states only near the Fermi wave number, $\pm \kf$, 
are taken into account,
 the Hamiltonian  is reduced to  
\begin{eqnarray}
& &{\cal H}
= \sum_{p k \sigma} p \vf k c^\dagger_{k,p,\sigma} c_{k,p,\sigma} 
+ \frac{1}{L} \sum_{p k k' q \sigma \sigma'} 
\nonumber \\
& &  
\Big\{
(g^0_{1\bot}\delta_{\sigma -\sigma'} + g^0_{1\Vert}\delta_{\sigma \sigma'})
c^\dagger_{k+q, p, \sigma} c^\dagger_{k'-q, -p, \sigma'} c_{k', p, \sigma'} c_{k, -p, \sigma}
\nonumber \\
& & +
(g^0_{2\bot}\delta_{\sigma -\sigma'} + g^0_{2\Vert}\delta_{\sigma \sigma'})
c^\dagger_{k+q, p, \sigma} c^\dagger_{k'-q, -p, \sigma'} c_{k', -p, \sigma'} c_{k, p, \sigma} 
\nonumber \\
& & + 
(g^0_{4\bot}\delta_{\sigma -\sigma'} + g^0_{4\Vert}\delta_{\sigma \sigma'})
c^\dagger_{k+q, p, \sigma} c^\dagger_{k'-q, p, \sigma'} c_{k', p, \sigma'} c_{k, p, \sigma} 
\Big\} \virg \nonumber \\
\end{eqnarray}
where $L = N_L a$ and 
  the energy dispersion 
is linearized at $\pm \kf$ as $\epsilon_{\pm\kf + k} - \mu \simeq \pm v_F k$
with $v_F = \sqrt{2} t a$.
Coupling constants are given by  
$g^0_{4 \bot} = g^0_{2 \bot} = Ua/2 + Va$, 
$g^0_{4 \Vert} = g^0_{2 \Vert} = Va$, 
$g^0_{1 \bot} = Ua/2$, and $g^0_{1 \Vert} = 0$.
 We note that 
  eq.(3) does not show the M-I transition 
     since there is no Umklapp scattering.

According to a theoretical suggestion by Schulz\cite{Schulz} 
and calculation of the mean-field theory\cite{Suzumura}, 
 Umklapp scattering in the case of quarter-filling 
appears through 
  the one-particle states with $\pm 3 \kf$, which are connected 
  to those of  the Fermi point  
    by $\pm 2 \kf$. 
 We take into account the states near
   $\pm 3 \kf$ systematically by the following procedure 
   in order to  derive the effective Hamiltonian, which consists of  
  the one-particle states only  near $\pm \kf$. 
  Representing $c_{k, p, \sigma}$ and $d_{k, p, \sigma}$ 
     in terms of Grassmann  algebra,  
     the partition function of  eqs.(1) and (2),  
   $Z$, is given    by 
   $Z = \int {\cal D}[d^*_{k, p, \sigma}d_{k, p, \sigma}]{\cal D}
     [c^*_{k, p, \sigma}c_{k, p, \sigma}]\e^{-S}$,  
 where 
 $S = S_0[c^*_{k, p, \sigma}c_{k, p, \sigma}] +
       S_0[d^*_{k, p, \sigma}d_{k, p, \sigma}]
          + \sum_{i = 0}^4 S_{{\rm int},i}$, and  
      $S_{{\rm int},i}$ denotes  the action corresponding to 
 ${\cal H}_{{\rm int},i}$. 
 After integrating $d_{k, p, \sigma}$, 
    the partition function, $Z$, is written as 
     $Z = Z^0_d \int {\cal D}[c^*_{k, p, \sigma}c_{k, p, \sigma}]
      \e^{-S_{eff}}$, 
 where 
   $Z^0_d = \int {\cal D}[d^*_{k, p, \sigma}d_{k, p, \sigma}]
     \e^{-S_0[d^*_{k, p, \sigma}d_{k, p, \sigma}]}$, and    
 $S_{eff} = S_0[c^*_{k, p, \sigma}c_{k, p, \sigma}] -
         \ln \lan \exp ( - \sum_{i = 0}^4 S_{{\rm int},i} )\ran_d$ 
($\lan \cdots \ran_d$ is the average by $S_0[d^*_{k, p, \sigma}d_{k, p, \sigma}]$). 
By calculating $S_{eff}$ perturbatively,
the effective Hamiltonian 
 is derived where 
    the diagram for the vertex includes   
 only the green functions of $d_{k, p, \sigma}$.
From the perturbation up to the second order of $U$ and/or  $V$,  
 the correction to the normal processes, $g^0_{i \Vert(\bot)}$, in eq.(3) 
 is calculated  as
\begin{eqnarray}
\Delta g_{1 \bot} 
   &=& - 4 D_1 \left(\frac{Ua}{2}\right) \left(\frac{Ua}{2} - Va \right) \virg \\
\Delta g_{2 \Vert} 
    &=& - 2 D_1 \left(Va\right)^2 \virg \\ 
\Delta g_{2 \bot} 
       &=& - 2 D_1 \left(\frac{Ua}{2}\right)^2 - 2 D_1 \left(\frac{Ua}{2} - Va \right)^2 
              \virg \\   
\Delta g_{4 \Vert} 
    &=& - 2 D_2 \left(Va\right)^2 \virg \\ 
\Delta g_{4 \bot} 
    &=& - 2 D_2 \left(\frac{Ua}{2} - Va \right)^2  
\virg
\end{eqnarray}
where $D_1 = (8 \pi t a)^{-1} \int^{\pi/2}_0 \d y (\sin y +
1/\sqrt{2})^{-1} = (8 \pi t a)^{-1} \sqrt{2} \ln (\sqrt{2}+1) 
  \simeq 1.25/(8\pi t a)$ and
$D_2 = (4 \pi t a)^{-1} \int^{\pi/2}_0 \d y (\sin y +
\cos y + \sqrt{2})^{-1} = (8 \pi t a)^{-1} 2 \sqrt{2}/(\sqrt{2}+1) \sim
1.17/(8 \pi t a)$.
 The $8\kf$-Umklapp scattering, which comes from the third order 
expansion, is obtained  as
\begin{eqnarray}
{\cal H}_{1/4} &=& \frac{(Ua)^2}{2 t^2}(Ua - 4Va) \nonumber \\ 
&\times& 
\sum_{p} \int \d x (\psi^\dagger_{p,+}\psi_{-p,+})^2 (\psi^\dagger_{p,-}\psi_{-p,-})^2
\nonumber \\ 
&+& \frac{(Ua)^3}{8 t^2} \sum_{p \sigma} \int \d x 
(\psi^\dagger_{p, \sigma}\psi_{-p, \sigma})^3 
\psi^\dagger_{p, -\sigma}\psi_{-p,-\sigma}  \virg
\end{eqnarray}
where the first term corresponds to the conventional 
  commensurability energy
\cite{Tsuchiizu} and 
$\psi_{p, \sigma} = 1/{\sqrt{L}} \sum_k \e^{\im k x} c_{k,p,
\sigma}$.  
 In  the third order, there are other contributions 
   leading to the correction to the normal processes,
 which   are disregarded in the following 
  due to  eqs.(4)-(8) being 
large enough for the present choice of parameters.

Here we utilize the bosonization method
\cite{Emery} 
 by introducing phase variables  for  the charge (spin) fluctuation,
  $\theta_{\rho}$ and  $\phi_\rho$, 
    ($\theta_{\sigma}$ and $\phi_\sigma$), which are defined by   
 $\theta_{\rho(\sigma)} = 
  \sum_q \pi \im /(q L) \e^{- \alpha |q|/2 -\im q x } 
    \sum_{k p \sigma} (\sigma) c^\dagger_{k+q, p, \sigma}c_{k, p, \sigma}$ 
and 
$\phi_{\rho(\sigma)} = 
\sum_q \pi \im /(q L) \e^{- \alpha |q|/2 -\im q x } 
\sum_{k p \sigma} (\sigma) p 
c^\dagger_{k+q, p, \sigma}c_{k, p, \sigma}$, respectively. 
The quantity $\alpha^{-1}$ is the ultraviolet cutoff. 
Those fields satisfy, 
$[\theta_{\nu}(x),\phi_{\nu'}(x')] 
= \im \pi {\rm sgn}(x-x') \delta_{\nu \nu'}$ 
($\nu = \rho$ or $\sigma$). 
The electron operator is expressed as 
 $\psi_{p, \sigma} = (2 \pi \alpha)^{-1/2}
     \exp\{ \im (p/2)(\theta_{\rho} + p \phi_{\rho} 
       + \sigma \theta_{\sigma} + \sigma p \phi_{\sigma})\} 
       \exp(\im \pi \Xi_{p \sigma})$ 
 with 
   $\Xi_{p +} = p/2 \sum_{p'}\N_{p'+}$ and  
  $\Xi_{p -} = p/2 \sum_{p'}\N_{p'-} +\sum_{p'}\N_{p' +}$
 where $\N_{p \sigma} = 
   \int {\rm d} x \psi_{p \sigma}^{\dagger} \psi_{p \sigma}$.  
In terms of these phase variables,
 the Hamiltonian is expressed as  
   ${\cal H} = {\cal H}_{\rho} + {\cal H}_{\sigma} + {\cal H}'$,  
   where 
\begin{eqnarray}
{\cal H}_{\rho} &=& \frac{v_{\rho}}{4 \pi} \int \d x
\left\{
\frac{1}{K_{\rho}} (\delx \theta_{\rho})^2 
+ K_{\rho} (\delx \phi_{\rho})^2 
\right\} \nonumber \\
&+& \frac{g_{1/4}}{2 (\pi \alpha)^2} \int \d x \cos 4 \theta_{\rho} \virg \\ 
{\cal H}_{\sigma} &=& \frac{v_{\sigma}}{4 \pi} \int \d x
\left\{
\frac{1}{K_{\sigma}} (\delx \theta_{\sigma})^2 
+ K_{\sigma} (\delx \phi_{\sigma})^2 
\right\} \nonumber \\
&+& \frac{g_{1\bot}}{(\pi \alpha)^2} \int \d x \cos 2 \theta_{\sigma} \virg \\
{\cal H}' &=& \frac{g'_{1/4}}{2 (\pi \alpha)^2} \int \d x \cos 4 \theta_{\rho} \cos 2 \theta_{\sigma} \point 
\end{eqnarray}
Here 
$K_\nu = \sqrt{B_\nu / A_\nu}$ and $v_\nu = \vf \sqrt{B_\nu  A_\nu}$
($\nu = \rho$ or $\sigma$),  
where $A_{\rho(\sigma)} = 1 + \{g_{4 \Vert}+(-)g_{4 \bot}+g_{2 \Vert}+(-)g_{2
\bot}-g_{1 \Vert}\}/(\pi \vf)$ 
and $B_{\rho(\sigma)} = 1 + \{g_{4 \Vert}+(-)g_{4 \bot}-g_{2 \Vert}-(+)g_{2
\bot}+g_{1 \Vert}\}/(\pi \vf)$ with 
$g_{i \Vert(\bot)} = g^0_{i \Vert(\bot)} + \Delta g_{i \Vert(\bot)}$. 
In eqs.(10)-(12), 
the coefficients of the nonlinear terms are respectively given 
by $g_{1\bot} = g^0_{1\bot} + \Delta g_{1\bot} = (Ua/2)\{1 - 4D_1(Ua/2 - 
Va) \}$,  
$g_{1/4} = (Ua)^2(Ua - 4Va)/(2 \pi \vf)^2 \times (a/\alpha)^2$
and $g'_{1/4} = (Ua)^3/(2 \pi \vf)^2 /2 \times (a/\alpha)^2$.
 For $V=0$,  the quantity $g_{1/4}$ is essentially 
the same as that reported by
 Schulz\cite{Schulz}
and Yonemitsu.\cite{Yonemitsu} 
A contribution from the fourth order of $U$ has also been suggested by 
Giamarchi\cite{Giamarchi}. 
The order parameters of the $2\kf$-SDW, $2\kf$-CDW and 
$4\kf$-CDW are given by
\begin{eqnarray}
O_{2\kf\mbox{-}\rm{SDW}} &=& \sum_{p\sigma} \sigma \e^{-\im 2 p \kf x}
\psi^\dagger_{p, \sigma} \psi_{-p, \sigma} \nonumber \\ 
&\propto& \sin(2 \kf x + \theta_{\rho}) \sin \theta_{\sigma} \virg \\
O_{2\kf\mbox{-}\rm{CDW}} &=& \sum_{p\sigma} \e^{-\im 2 p \kf x}
\psi^\dagger_{p, \sigma} \psi_{-p, \sigma} \nonumber \\ 
&\propto& \cos(2 \kf x + \theta_{\rho}) \cos \theta_{\sigma} \virg \\
O_{4\kf\mbox{-}\rm{CDW}} &=& \sum_{p} \e^{-\im 4 p \kf x}
\psi^\dagger_{p, +} \psi^\dagger_{p, -}
\psi_{-p, -} \psi_{-p, +} \nonumber \\
&\propto& \cos(4 \kf x + 2 \theta_{\rho}) \point
\end{eqnarray}

We investigate the possible states in the limit of low energy.   
 Since the scaling dimension of ${\cal H}'$  given by 
 $2 - 8 K_\rho - 2K_\sigma$  is  smaller than 
   that of the other nonlinear terms, 
  the term may be safely neglected for determining 
    the phase diagram. 
 In this case, the Hamiltonian is divided into 
    the charge part and spin part. 
For the spin part, 
 as long as $g_{1 \bot} > 0$, 
  the quantities, $g_{1\bot}$ and $K_{\sigma}$ tend toward 0 and 1,
    respectively, for the low energy limit, due to $SU(2)$ symmetry,  
        and the excitation is gapless. 
Then the long range  correlation functions  are given as 
 $\lan \sin \theta_{\sigma}(x) \sin \theta_{\sigma}(0)\ran
   \sim x^{-1}\ln^{1/2}(x)$ and    
$\lan \cos \theta_{\sigma}(x) \cos \theta_{\sigma}(0)\ran
\sim x^{-1}\ln^{-3/2}(x)$.\cite{Giamarchi-Schulz}
 The low energy property for the charge part 
is determined by the following RG equations,  
\begin{eqnarray}
\frac{d}{d l} K_{\rho}(l) &=& - 8 G_{1/4}^2(l) K_{\rho}^2(l) \virg \\
\frac{d}{d l} G_{1/4}(l) &=& \left[2 - 8 K_{\rho}(l)\right] G_{1/4}(l) \virg
\end{eqnarray}  
with the initial conditions  given by $K_{\rho}(0) = K_{\rho}$ 
and $G_{1/4}(0) = g_{1/4}/(2 \pi v_{\rho})$.
Here $l = \ln (\alpha'/\alpha)$ with the new length scale 
 $\alpha'$ larger than $\alpha$.   
The solution of the above RG equations is obtained  by
\begin{equation}
G^2_{1/4}(l) - 1/(2K_{\rho}(l)) - 2 \ln K_{\rho}(l) = const \virg
\end{equation}
with a decrease of $K_{\rho}$ with increasing $l$.

\begin{figure}[t]
\hspace*{0.5cm}
\epsfxsize=7.5cm
   \epsffile{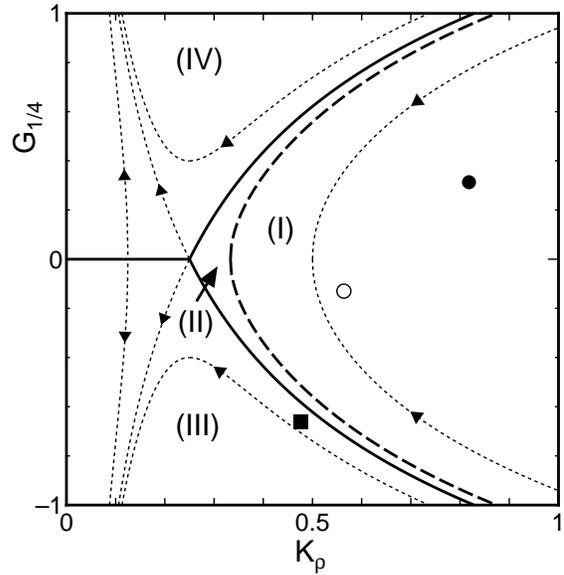}
\vspace*{-.5cm}
\caption{ 
Scaling flows ( dotted curves ) derived from eqs.  (16) and (17)
 on the plane of $K_\rho$ and $G_{1/4}$,  
 where the solid and dashed curves denote their boundaries.    
The parameters in regions  (I) and (II) lead to weak coupling,    
 where $K_{\rho}(\infty) = 1/3$ on the dashed curve.  
 A strong coupling regime is found  
 in both regions, (III) and (IV).
The closed  circle, open  circle and closed square  
correspond to the initial values 
 of $V/t =$ 0, 2 and 6, respectively, 
 with the fixed  $U/t = 5$.     
}
\end{figure}
The RG flows are shown in Fig.1,  
 where the value of $\alpha \simeq 2 a / \pi$ is used.\cite{Suzumura-II}
In the regions (I) and (II), 
  the nonlinear term becomes irrelevant, indicating a metallic state.
 The correlation functions with  long distance are given by 
  $\lan \sin \theta_{\rho}(x) \sin \theta_{\rho}(0)\ran
   \sim \lan \cos \theta_{\rho}(x) \cos \theta_{\rho}(0)\ran
    \sim x^{- K_{\rho}(\infty)}$
and $\lan \cos 2 \theta_{\rho}(x) \cos 2 \theta_{\rho}(0)\ran
   \sim x^{- 4 K_{\rho}(\infty)}$. 
Therefore the dominant state exhibits a crossover 
 between $2\kf$-SDW and $4\kf$-CDW
  at $K_{\rho}(\infty) = 1/3$,\cite{Schulz-II} where 
 $K_{\rho}(\infty)$ is determined by  
$G^2_{1/4}(0) - 1/(2K_{\rho}(0)) - 2 \ln K_{\rho}(0) =  
- 1/(2K_{\rho}(\infty)) - 2 \ln K_{\rho}(\infty)$. 
The phase boundary determined by  $K_{\rho}(\infty) = 1/3$ 
  is shown by the dashed curve.
Both  regions, (III) and (IV) are insulating states where 
 $G_{1/4} \to - \infty$ and $\theta_{\rho}$ is locked as $0, \pi/2 $ 
 mod $\pi$ in (III), 
and $G_{1/4} \to \infty$ and $\theta_{\rho} = \pm \pi/4$ mod $\pi$ in
(IV).
The ordered state of $4\kf$-CDW leading to charge ordering 
 is realized in (III) 
   and $\theta_{\rho} = 0$ ($\theta_{\rho} = \pi/2$)
    corresponds to the state in which 
     the charge is rich  at  the even (odd) sites. 
The closed  circle, open  circle and closed  square  
correspond to the initial condition 
 of $V/t =$ 0, 2 and 6, respectively, with  $U/t = 5$, 
showing that  the metallic state with  $2\kf$-SDW   in (I) moves   to  
 the insulating state with the $4\kf$-CDW  ordering 
   with increasing $V$.

\begin{figure}[t]
\hspace*{0.5cm}
\epsfxsize=7.5cm
   \epsffile{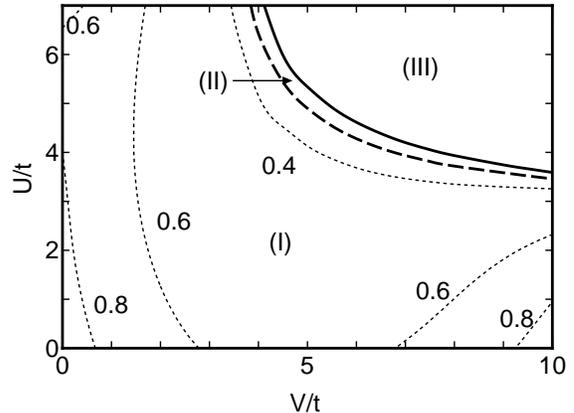}
\vspace*{-.5cm}
\caption{
The phase diagram on the plane of $V/t$ and $U/t$,  
 where regions (I), (II) and  (III) 
  separated by  boundaries (solid and dashed curves)
    correspond to those in Fig. 1.   
   The metallic states with   
   $2\kf$-SDW and $4\kf$-CDW are obtained in regions (I) and (II),
  respectively, 
   while region (III) is the insulating state with 
 $4\kf$ charge ordering. 
 The dotted  curves in the metallic state denote $K_\rho(\infty)$.
}
\end{figure}
The phase diagram on the plane of $U/t$ and $V/t$ 
 is shown in Fig.2.
The insulating phase 
appears for a large strength of $U$ and $V$.
In addition, 
for large $V$ and small $U$, 
$K_{\rho}(\infty)$ approaches  unity indicating that 
 the superconducting fluctuation is enhanced in this region. 
Figure 2 is qualitatively the same as that  derived from 
the numerical diagonalization.
\cite{Penc-Mila-I,Mila-Zotos,Sano-Ono,Nakamura-Kitazawa-Nomura}
In the insulating phase, which corresponds to the region (III) in Fig.1, 
the $4\kf$-CDW, \ie charge disproportionation, is realized. 
The state is consistent with the mean-field result.\cite{Seo-Fukuyama}
Thus, the insulating state obtained by the numerical diagonalization 
 is due to the formation of charge disproportionation,  which 
 originates in both  $U$ and $V$ interactions.        
Here, we discuss the effects of the nonlinear term, 
$\cos 4 \theta_{\rho} \cos 2 \theta_{\sigma}$ of eq.(12), 
 which may give minor changes on the phase boundary. 
In the insulating region, 
 the term seems to give the correction of $g_{1 \bot}$
as $g_{1 \bot} \to g_{1 \bot} + g'_{1/4} \lan \cos 4 \theta_{\rho} 
\ran / 2$ with $\lan \cos 4 \theta_{\rho} \ran > 0$. 
Such a coupling constant gives rise to the spin gap. 
However, the procedure breaks the $SU(2)$ symmetry obviously. 
Therefore, it is expected that the third order correction of $K_{\sigma}$, which has been
discarded in the present treatment, would restore the symmetry, and 
the spin excitation remains gapless in the insulating region.     

Within the present analysis, 
the insulating state exhibits the fluctuation of 
$2\kf$-SDW,
while $4\kf$-CDW exists as the true long range order. 
When three dimensionality is introduced by the interchain coupling, 
the fluctuation of $2\kf$-SDW becomes  the true order. 
However, the characteristic temperature of $2\kf$-SDW 
is expected to be smaller than that of $4\kf$-CDW. 
This conclusion may be relevant to 
  the following experimental
    observation in (DI-DCNQI)$_2$Ag.
The magnetic order of $2\kf$-SDW is observed at 5.5 K,
whereas the resistivity shows insulating behavior 
at least below room temperature.
\cite{Hiraki-Kanoda-II}  
In addition, charge disproportionation  is observed below 220 K.
\cite{Hiraki-Kanoda}  

Finally, the effects of the next-nearest-neighbor repulsion, $V_2$, 
 are briefly discussed. 
Up to the second order perturbation in the present formulation, 
 the properties of the charge fluctuation do not show a qualitative
   change.  
On the other hand,  
 the coefficient of the nonlinear term of the spin fluctuation, 
 $g_{1 \bot}$,  
  is changed as  
   $(Ua/2 - V_2 a)\{1 - 4D_1(Ua/2 - Va + V_2 a)\}$.
Since the nonlinear term  with  positive (negative)  $g_{1 \bot} $
 favors  $2\kf$-SDW ($2\kf$-CDW), 
 it is obvious that 
    the repulsive next-nearest-neighbor interaction 
        stabilizes $2\kf$-CDW. 
In fact, 
 the spin excitation becomes gapped and $2\kf$-CDW is realized 
   when $U/2 < V_2$, {\it independent} of the strength of $V$ 
 as seen in the mean-field result.
\cite{Tomio}
Thus the quantity $V_2$ stabilizes (suppresses)
 $2\kf$-CDW ($2\kf$-SDW). 
However, 
the coexistence of two kinds of $2\kf$ density waves, 
which is predicted by the mean-field theory,
\cite{Kobayashi-Ogata-Yonemitsu} 
cannot be understood from such a calculation, which needs  
 further exploration.    

In conclusion, we investigated both normal scattering and Umklapp  scattering 
 analytically for  one-dimensional quarter-filled 
  electron systems with on-site and nearest-neighbor interactions. 
 We have shown that the increase of both $U$ and $V$ leads  to 
 the insulating state with charge disproportionation  
 and that the next-nearest-neighbor interaction  
  suppresses (enhances) the fluctuation of $2\kf$-SDW ($2\kf$-CDW).

\section*{Acknowledgment}
The authors would like to thank H. Fukuyama for his useful comments.
 This work was supported by 
 a Grant-in-Aid 
 for Scientific  Research  from the Ministry of Education, 
Science, Sports and Culture (No.11740196)
 and by  
a Grant-in-Aid 
 for Scientific  Research  from the Ministry of Education, 
Science, Sports and Culture (No.09640429).


\begin{thebibliography}{99}
\bibitem{Suzumura-II}
This relation can be derived by comparing the response functions near
 $q = 2\kf$ in the non-interacting case, which are  
calculated by the conventional fermionic method and by the bosonization method.
\end{thebibliography}
\end{document}